\documentclass[12pt]{article}
\usepackage{amssymb}
\usepackage{graphicx}

\newcommand{\be}{\begin{equation}}
\newcommand{\ee}{\end{equation}}
\newcommand{\bea}{\begin{eqnarray}}
\newcommand{\eea}{\end{eqnarray}}

\renewcommand{\em}{\it}

\renewcommand{\title}[1]{\vbox{\center\LARGE
\bf\mathversion{bold}{#1}\mathversion{normal}
}\vspace{5mm}}
\renewcommand{\author}[1]{\vbox{\center#1}\vspace{5mm}}
\newcommand{\address}[1]{\vbox{\center\em #1}}
\newcommand{\email}[1]{\vbox{\center\tt#1}\vspace{5mm}}

\newcommand{\bref}[1]{(\ref{#1})}

\begin{document}

\begin{titlepage}
\begin{center}
\vspace{5mm}
\hfill {\tt HU-EP-10/11}\\
\vspace{40mm} 
{\font\titlerm=cmr10 scaled\magstep4 {\titlerm 
Gauge Theory Wilson Loops and \\ Conformal Toda Field Theory}}

\bigskip

\author{\large Filippo Passerini}
\address{Institut f\"ur Physik, Humboldt-Universit\"at zu Berlin,\\
Newtonstra{\ss}e 15, D-12489 Berlin, Germany}

\email{filippo@physik.hu-berlin.de}

\end{center}

\abstract{
\noindent
The partition function of a family of four dimensional ${\cal N}=2$  gauge theories   has been recently  related to correlation functions of two dimensional conformal Toda field theories.  For  $SU(2)$ gauge theories,   the associated two dimensional theory is  $A_1$ conformal    Toda field theory,  i.e.   Liouville  theory.  For this case the relation has  been extended showing   that the  expectation value of gauge theory   loop operators  can be reproduced in Liouville  theory  inserting in the correlators  the monodromy  of  chiral degenerate fields.  In this  paper we  study Wilson loops  in  $SU(N)$ gauge theories in the  fundamental and  anti-fundamental representation of the gauge  group  and   show  that  they are  associated to   monodromies of a certain chiral degenerate operator of $A_{N-1}$ Toda field theory.  The  orientation of the curve along  which the monodromy is  evaluated selects between   fundamental and  anti-fundamental  representation.  The analysis  is performed using properties of the monodromy group of the generalized hypergeometric equation, the differential equation satisfied by a class of  four point  functions relevant for  our  computation.           
}

\vfill

\end{titlepage}

%%%%%%%%%%%%%%%%%%%%%%%

\section{Introduction and Discussion}

In a recent paper \cite{Gaiotto:2009we},   Gaiotto has  constructed a large class  of  four dimensional ${\cal N}=2$ gauge theories that describe the low energy dynamics of a stack of $N$ M5-branes compactified on a punctured Riemann surface $C_{(f_a),g}$.  The surface is characterized by the  genus $g$  and the number of punctures $(f_a)$.  There are different types of  puncture and each type is labeled by a Young  tableaux with $N$ boxes.   Following the  construction in  \cite{Gaiotto:2009we},   it is possible to associate   to any  surface $C_{(f_a),g}$   a  four dimensional  gauge  theory ${\cal T}_{(f_a),g}$    characterized  by  the  same  data   labeling the surface.   The  different S-duality  frames   of   the  gauge  theory correspond to  the  different ways of  sewing  the Riemann  surface  from  pairs  of  pants.

In \cite{Alday:2009aq},  Alday, Gaiotto and  Tachikawa  (AGT)  have related the four dimensional theory  ${\cal T}_{(f,g)}$ when $N=2$  to   two dimensional Liouville theory defined  on the Riemann surface  $C_{(f,g)}$\footnote{In the $N=2$ case, there is only one type on puncture and the gauge theory includes only $SU(2)$ gauge groups and  $SU(2)$ flavor groups.}.  An important ingredient of the   analysis  is given  by the the Nekrasov partition function $Z_\textmd{\tiny Nekrasov}$ \cite{Nekrasov:2002qd}\cite{Nekrasov:2003rj},  that is computed considering a  two parameters  deformation of the gauge theory and  is written as $Z_\textmd{\tiny Nekrasov}=Z_\textmd{\tiny classical}Z_\textmd{\tiny 1-loop}Z_\textmd{\tiny instanton}$, where the three factors are the classical, the quantum and the instanton contributions.  Given a certain S-duality frame,  $Z_\textmd{\tiny instanton}$ is identified with the BPZ \cite{Belavin:1984vu} conformal block ${\cal F}_\alpha^\beta$  associated to the pants decomposition correspondending to the chosen S-duality frame.   The $Z_\textmd{\tiny instanton}(\epsilon,a,m)$    depends on the deformation parameters $(\epsilon_1,\epsilon_2)$, the Coulomb branch coordinates $a$ and the mass parameters $m$. These  quantities   are related respectively to the  Liouville coupling constant $b$ and the momenta of the internal states $\alpha$  and external states $\beta$\footnote{See \cite{Alday:2009aq} for the precise dictionary.}.  

Using the results of   Pestun  \cite{Pestun:2007rz}, the partition function of the gauge theory defined on  $S^4$ was identified with a Liouville correlation function, considering Liouville theory  on  the Riemann surface $C_{f,g}$ with coupling constant $b=1$.  It results 
\bea\label{cfunct}
Z_{{\cal T}_{f,g}}=\int [da] Z_\textmd{\tiny Nekrasov}\bar{Z}_\textmd{\tiny Nekrasov}=\langle V_{\beta_1}\ldots V_{\beta_f}\rangle
\eea   
where  $V_{\beta_1}\ldots V_{\beta_f}$ are Liouville  primary fields associated to the $f$ punctures of  $C_{f,g}$ with momenta     related to the  mass parameters of the gauge theory.  The relation \bref{cfunct} is  satisfied  because, besides  the  correspondence  between  the conformal  blocks and $Z_\textmd{\tiny instanton}$ that we have   already  mentioned,  it  is  possible  to  show  that the factors $Z_\textmd{\tiny classical}$ and $Z_\textmd{\tiny 1-loop}$ of the $Z_\textmd{\tiny Nekrasov}$  reproduce the product of three  point  functions of  Liouville primary  fields given by the DOZZ formula \cite{Dorn:1994xn}\cite{Zamolodchikov:1995aa}\cite{Teschner:1995yf}.

The AGT proposal was  soon  extended  by  Wyllard \cite{Wyllard:2009hg}  to the $N>2$ case.  In this  configuration  the Liouville  theory is  replaced  by the   conformal $A_{N-1}$ Toda field theory\footnote{See for  instance \cite{Fateev:2007ab}. We quickly review Toda field theory in the next section.}, a generalization  of Liouville theory that is invariant under the ${\cal W}_{N}$ algebra, an extension of the Virasoro algebra.    Given a generic punctured surface,  the partition function of the corresponding  gauge theory   is given by a correlation function of  the Toda primary fields associated to the punctures.  Since in the  $N>2$  case  there are many types of punctures, it is necessary to consider different types of primaries   \cite{Wyllard:2009hg}\cite{Kanno:2009ga}.   Despite the  fact that  Liouville theory is a $A_1$ Toda theory, the generalization from $N=2$ to $N>2$  is not  straightforward because  the higher  rank  Toda  theories  are  much  less  understood  than  Liouville  theory.   First,  the  three  point  function of  primary  fields  is  known exactly only  when one  of  the  insertion is a certain degenerate field \cite{Fateev:2005gs}.  Furthermore, differently from Liouville ($A_1$ Toda), in the general case  it  is   not possible  to decompose the higher point  correlation  functions in terms  of  three  point  functions of ${\cal W}_{N}$ primary fields and ${\cal W}_{N}$ conformal blocks \cite{Bowcock:1993wq}.    Due to these complications,  it  is  not  possible to  write a differential equation  for  the four point  function with only one degenerate insertion, like for the Liouville  case. It  is  however possible  to write a differential equation  for  a  four point  function with a degenerate  and a  semidegenerate field \cite{Fateev:2005gs}.    Other recent results  concerning the  relation between Toda theories and gauge theories    are given in \cite{Mironov:2009by}\cite{Bonelli:2009zp}\cite{Alday:2009qq}\cite{Nanopoulos:2009uw}\cite{Dijkgraaf:2009pc}\cite{Itoyama:2009sc}\cite{Eguchi:2009gf}\cite{Schiappa:2009cc}\cite{Mironov:2009ib}\cite{Fujita:2009gf}\cite{Mironov:2010zs}.

In \cite{Pestun:2007rz}, Pestun considers also the case where a circular supesymmertic Wilson loop is inserted  in the partition function.  It results that the  expectation value  of the supersymmetric circular  Wilson loop in  ${\cal N}=2$ gauge theories is    given by the  following  matrix  model\footnote{In \cite{Pestun:2007rz}, Pestun derives  also the  matrix  model for the supersymmetric circular  Wilson loop in ${\cal N}=4$ theory, proving  the conjectures of  \cite{Erickson:2000af} \cite{Drukker:2000rr}.}  
\bea\label{wlpest}
\langle W_R\rangle=\int [da] Z_\textmd{\tiny Nekrasov}\bar{Z}_\textmd{\tiny Nekrasov}  \hbox{Tr}_R\, e^{2\pi i  a },
\eea
where $R$ is a   representation  of the  gauge group\footnote{Other aspects  of Wilson loops in ${\cal N}=2$ theories have been recently analyzed  in \cite{Rey:2010ry}.}.  The Liouville theory description of line  operators for  theories with $SU(2)$ gauge  groups was analyzed in  \cite{Drukker:2009id} \cite{Alday:2009fs}.    The authors show that the  insertion of  a supersymmetric  loop operator in the partition function of a ${\cal N}=2$ gauge theory  correspond in the Liouville theory  to the  monodromy of a chiral degenerate  operator.  With this prescription t'Hooft operators, dyonic operators  and  Wilson operators are  treated  in the  same  framework and  the  nature of the operators is  completely  encoded  by  the  curve on $C_{f,g}$ along which the monodromy is  computed.  A precise dictionary between   charges of  loop  operators and unoriented closed curves on $C_{f,g}$  was given in \cite{Drukker:2009tz}.

In this paper we study the two dimensional realization of  circular supersymmetric Wilson loops in $SU(N)$ gauge theories with $N>2$. In particular, we  generalize the proposal of  \cite{Drukker:2009id}\cite{Alday:2009fs}   showing that the fundamental and anti-fundamantal  Wilson loops in  $SU(N)$ ${\cal T}_{(f_a),g}$  theories\footnote{For these theories $SU(N)$ is the gauge group  with the  highest rank but  lower  rank  groups are  also  admitted in the  quiver.}  are  associated to monodromies  of  a particular chiral degenerate  operators of  $A_{N-1}$ Toda  field  theory. The  analysis  is  performed  using  properties  of  the  monodromy  group of the  generalized  hypergeometric  equation, that  is  the differential  equation satisfied  by  Toda  four  point  functions with two degenerate fields, i.e. the class of correlation function relevant for our computation.  Our  result  is in  agreement  with Pestun's results  \bref{wlpest} when we  set to one the  Toda coupling constant.    We show that  Wilson loops in the  fundamantal and  anti-fundamantal  representation of an $SU(N)$ gauge  group are obtained  from the same degenerate operator evaluating the  monodromy  along     curves   with  opposite  orientation.   This  implies that   loop  operators  in $SU(N)$ ${\cal T}_{(f_a),g}$ gauge theories are  associated to    oriented curves on the Riemann surface $C_{(f_a),g}$.  The   orientation for the curves  is  a new  features of the  $N>2$  case, since for the $SU(2)$ gauge theories  the   orientation of the monodromy is not relevant because  the  representations of this group are  real. 

The rest of the paper is  organized as  follows. In section 2 we  review $A_{N-1}$ conformal Toda field  theory, focusing  in  particular on the four point function with two degenerate  insertions. We describe the generalized hypergeometric  differential equation satisfied by this correlation function and review basic  properties of its  monodromy group.  In section 3 we show  that  $SU(N)$ Wilson loops in the fundamental and anti-fundamantal representations   are  associated  to  monodromies of  a particular chiral degenerate Toda  field.   Appendix A reviews basic  properties of Lie  algebras and  Appendix B presents  an  explicit  representation of the hypergeometric monodromy group.   

\textit{Note:} The Toda field theory description of $SU(N)$ Wilson loops in any representation of the gauge group has been recently obtained in  \cite{Drukker:2010jp} using a different approach. We thank the authors for informing us  of their results before publication.

\section{$A_{N-1}$ Conformal Toda Field Theory}

We collect  in this  section some  basic  and  known  facts  about  $A_{N-1}$  Toda field  theory,  following mostly  \cite{Fateev:2007ab}.   The dynamical  fields in the  theory  are  a  set  of  $N-1$  scalars $\varphi_i$ ($i=1,\ldots ,N-1$)  that   propagate  on  a  two  dimensional Riemann surface. The   set  of  scalars  form the components  of  an $N-1$ vector $\varphi$ defined in the root  space  of  the  $A_{N-1}$  Lie algebra, i.e. $\varphi=\sum_{k=1}^{N-1}\varphi_{k}e_k$ where $e_k$ is  a  simple  root\footnote{We review few aspects of Lie algebra theory in Appendix A.}. The  action is   
\begin{equation}\label{todaf}
S_{A_{N-1}}=\int dx^2 \sqrt{g}\left(\frac{1}{8\pi}g^{\alpha\beta}\langle\partial_\alpha\varphi,\partial_\beta\varphi\rangle+\frac{\langle Q,\varphi\rangle}{4\pi}R+\mu\sum_{k=1}^{N-1}e^{b\langle e_k,\varphi\rangle} \right)
\end{equation}
where   $b$  is  the  dimensionless  coupling  constant, $\mu$ is a constant called  cosmological  constant,  $g_{\alpha\beta}$ and $R$ are  the non-dynamical metric and  curvature of the Riemann surface. The scalar  product in the  root  space  is  defined  such  that  $\langle e_i,e_j\rangle=K_{ij}$  where  $K_{ij}$ is  the  Cartan  matrix of the $A_{N-1}$ algebra \bref{carmat}.  Conformal invariance of the theory requires the  background  charge  $Q$ to be  related to the coupling constant $b$   as
\begin{equation}
Q=q\rho
\end{equation}   
where $q=\left(b+\frac{1}{b}\right)$ and  $\rho$ is the Weyl vector.  In the  following we  will consider $g_{\alpha\beta}=\delta_{\alpha\beta}$ and  we  will use the complex notation for the  two dimensional coordinates, i.e.  $z=x_1+ix_2$ and $\bar{z}=x_1-ix_2$.

Besides  conformal invariance,    $A_{N-1}$ conformal Toda field theory  enjoys also  higher spin symmetries.  In total  there are $N-1$ holomorphic currents $W^{(i+1)}$ ($i=1,\dots,N-1$) with conformal dimension  $(i+1)$, where    $W^{(2)}=T$ is the usual stress tensor with conformal dimension $2$.  The $N-1$ symmetry  currents form a ${\cal W}_{N}$ algebra, that   is a  consistent  extension of the  Virasoro symmetry.  The Laurent expansions of the currents are  
\bea
W^{(i+1)}(z)=\sum_{n}\frac{W^{(i+1)}_n}{z^{n+i+1}}
\eea
and the  system includes  also antiholomorphic currents with analogous properties so that  the  total symmetry  is  ${\cal W}_{N}\times\overline{{\cal W}}_{N}$,  i.e. a product of an  holomorphic and  an  anti-holomorphic  ${\cal W}$-algebra.      We note that when $N=2$, the  only holomorphic current is the stress tensor, thus the  ${\cal W}_{2}$ algebra is the Virasoro  algebra. Indeed the   $A_{1}$ Toda field theory is  the  well studied Liouville theory,  a theory  that  posses  Virasoro invariance.        

The  ${\cal W}_{N}$ primary  fields $V$, are defined such that 
\begin{equation}
W^{(i+1)}_0V=w^{(i+1)}V,\qquad W^{(i+1)}_nV=0\qquad\hbox{for}\qquad n>0
\end{equation}
and like for the Virasoro algebra, the descendant fields are  obtained acting on the primaries  with the operators $W^{(i+1)}_{-n}$ where  $n>0$.
In Toda  field  theory  the primaries  are realized as exponential fields parameterized by $\alpha$, a vector in the root space of the $A_{N-1}$ algebra
\begin{equation}\label{pri}
V_\alpha=e^{\langle \alpha, \varphi \rangle}
\end{equation}
and their  conformal dimension  $\Delta(\alpha)\equiv w^{(2)}(\alpha)$ is  given  by 
\begin{equation}\label{condim}
\Delta(\alpha)\equiv w^{(2)}(\alpha)=\frac{1}{2}\langle \alpha, 2 Q -\alpha \rangle.
\end{equation}
An important set of  primaries  is  given by the completely degenerate fields. For these fields, the vector $\alpha$ is given by 
\begin{equation}\label{deg}
\alpha=-b\Omega_1-\frac{1}{b}\Omega_2
\end{equation}
where $\Omega_1$  and $\Omega_2$ are  two highest weights   of  finite  dimensional representations of the algebra  $A_{N-1}$.  In the OPE of the degenerate fields with a generic primary   appear only a finite set of primaries \cite{Fateev:1987zh}  
\begin{equation}\label{opedeg}
V_{-b\Omega_1-\frac{1}{b}\Omega_2}\cdot V_{\alpha}=\sum_{s,t}\,[V_{\alpha'_{s,t}}]
\end{equation}
where $\alpha'_{s,t}=\alpha-bh_s^{\Omega_1}-\frac{1}{b}h_t^{\Omega_2}$ and $h_s^{\Omega}$ are the weights of the representation of $A_{N-1}$ that has $\Omega$ as highest weight.  $[V_{\alpha'_{s,t}}]$ represent the family of operators that are descendants of $V_{\alpha'_{s,t}}$. We omit numerical factors on the right hand  side of the formula \bref{opedeg}.

\subsection{Four Point Correlation  Function}
 
In \cite{Fateev:2005gs}   the  four point   correlation  function with two  degenerate  insertions $V_{-b\omega_1}$ and   $V_{-b \omega_{N-1}}$   has  been  computed\footnote{The authors analyzed a  more general configuration with one degenerate and one semidegenerate insertion. We  specify to the two degenerate  insertions case because this is the configuration that we will need in the following.}.   It results
\bea
\langle V_{\alpha_1}(0) V_{-b\omega_1}(z,\bar{z}) V_{-b\omega_{N-1}}(1) V_{\alpha_2}(\infty) \rangle=|z|^{2b\langle \alpha_1, h_1 \rangle}|1-z|^{\frac{-2b^2}{N}}G(z,\bar{z}) 
\eea 
where  $G(z,\bar{z})$ satisfies the  generalized hypergeometric  differential  equation in each of the  two  complex  variables $z$ and $\bar{z}$.  In details
 \bea\label{hypeq}
D(A_1,\ldots,A_{N};B_1,\ldots,B_{N}) G(z,\bar{z})=0\\
 \bar{D}(A_1,\ldots,A_{N};B_1,\ldots,B_{N}) G(z,\bar{z})=0
\eea 
where
\bea
D(A_1,\ldots,A_{N};B_1,\ldots,B_{N})=z(z\partial+A_1)\dots(z\partial+A_N)-(z\partial+B_1-1)\ldots(z\partial+B_N-1)\nonumber\\
\eea
and $\partial=\frac{\partial}{\partial z}$. The  parameters $A_k$  and $B_k$ are related to the  Toda momenta  as 
\bea\label{momenta}
A_k=-b^2+b\langle\alpha_1-Q,h_1\rangle+b\langle\alpha_2-Q,h_k\rangle\nonumber\\
B_k=1+b\langle\alpha_1-Q,h_1\rangle-b\langle\alpha_1-Q,h_{k+1}\rangle
\eea
where  we  take $h_{N+1}=h_{1}$ so that  $B_N=1$.  $h_k$ are the $N$ weights of the fundamental representation of the $A_{N-1}$ algebra\footnote{See Appendix A, formula  \bref{fwei}.}. The $\bar{D}$ operator is obtained from  $D$ replacing $z$ with $\bar{z}$.  The solutions of the differential equation \bref{hypeq} are defined on the Riemann sphere  and  have  three  singular  points, namely $0,1,\infty$, that are the positions  where we  have  located three of the  fields. In each punctured neighborhood of the  singularities is possible to define   $N$ linearly independent solutions. We denote these  solutions  as 
\bea
\Lambda^{(s)}&=&(\Lambda^{(s)}_1,\ldots,\Lambda^{(s)}_N)\qquad \mbox{defined in a neighborhood of $0$,}\nonumber\\
\Lambda^{(t)}&=&(\Lambda^{(t)}_1,\ldots,\Lambda^{(t)}_N)\qquad\mbox{defined in a neighborhood of $1$,}\nonumber\\
\Lambda^{(u)}&=&(\Lambda^{(u)}_1,\ldots,\Lambda^{(u)}_N)\qquad\mbox{defined in a neighborhood of $\infty$.}\label{hsol}
\eea
For an explicit expression of these functions, see  for  instance \cite{norlund}. Through analytical continuation it is  possible to extend these solutions outside  their domain of definition and  it is thus  possible   to  consider analytical continuations  along closed paths.  If  the closed path  encircles one or more singularities,   the vector of  solution is linearly  transformed by an element of $GL(N,\mathbb{C})$, i.e. a monodromy matrix.   Given a  vector of $N$ linearly independent solutions, it is possible to compute the  monodromy matrices around all the three singularities.  These matrices,  denoted as  $M_{(0)}$, $M_{(1)}$ and $M{(\infty)}$,  represent the monodromies computed  around  the  three  homotopy  classes  of the  three  punctured  sphere.     The monodromy matrices form a subgroup of $GL(N,\mathbb{C})$, the monodromy group, defined by the following relation
\bea\label{monrel}
M_{(\infty)}M_{(1)}M_{(0)}=1.
\eea
This  group is a representation on the linear space  of  solutions  of  the first homotopy  group  of the Riemann sphere  with  three  punctures. We note that the monodromy group is invariant under conjugation inside $GL(N,\mathbb{C})$. That is, given three  matrices satisfying the relation \bref{monrel} and given $X\in GL(N,\mathbb{C})$, also the conjugated matrices $\tilde{M}=X^{}MX^{-1}$ satisfies the relation \bref{monrel}. The conjugation  relates monodromy matrices  that are  associated  to  vector  solutions  related  by the  linear  transformation $\tilde{\Lambda}=X \Lambda$.  This  implies that  given  a  representation of  the  group  for  a  certain basis of  independent  solutions, it  is  possible to know  the representation of  the  group  for  a  different set  of  solutions through a  simple  conjugation.  It results that the three vector  solutions  in \bref{hsol} have diagonal  monodromy matrices  respect to the  singularity   where  they are nearby defined. 

The  conformal blocks in the   $s,t,u$ channel   are  given by 
\bea\label{cblock}
{\cal F}^{(s)}_k=z^{b\langle \alpha_1, h_1 \rangle}(1-z)^{\frac{-b^2}{N}}\Lambda^{(s)}_k,\nonumber\\
{\cal F}^{(t)}_k=z^{b\langle \alpha_1, h_1 \rangle}(1-z)^{\frac{-b^2}{N}}\Lambda^{(t)}_k,\nonumber\\
{\cal F}^{(u)}_k=z^{b\langle \alpha_1, h_1 \rangle}(1-z)^{\frac{-b^2}{N}}\Lambda^{(u)}_k
\eea
where $k=1,\ldots,N$.  The  four point  function is obtained considering  bilinear combinations  of ${\cal F}(z)$ and  $\bar{\cal F}(\bar{z})$ that give a single valued function. More  precisely 
\bea\label{eqt}
\langle V_{\alpha_1}(0) V_{-b\omega_1}(z,\bar{z}) &&V_{-b\omega_{N-1}}(1) V_{\alpha_2}(\infty) \rangle=\\
&&\sum_{k,r}C_{kr}^{(s)}{\cal F}^{(s)}_k\bar{{\cal F}}^{(s)}_r=\sum_{k,r}C_{kr}^{(t)}{\cal F}^{(t)}_k\bar{{\cal F}}^{(t)}_r=\sum_{k,r}C_{kr}^{(u)}{\cal F}^{(u)}_k\bar{{\cal F}}^{(u)}_r\nonumber
\eea  
where $C_{kr}^{(s,t,u)}$ are diagonal matrices whose  entries are related  to the  three  point  functions in the $s,t,u$-channel\footnote{This relation was  used  in \cite{Teschner:1995yf}\cite{Fateev:2007ab} to obtain equations for three  point  correlation  functions.}.

\section{Wilson Loops in Conformal Toda Field Theory}

\begin{figure}[h]
\center{\includegraphics[width=8cm]{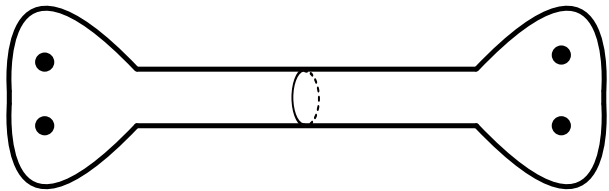}}
\caption{A sphere with four punctures in a given pair of pants decomposition. The tube  in  between the  two  couples of  punctures  is  associated to a gauge group and  the closed  curve is  associated to a Wilson loop operator.}\label{figure1}
\end{figure}

In \cite{Drukker:2009id} \cite{Alday:2009fs}, $SU(2)$ gauge theory loop operators  have  been  associated  to monodromies of chiral degenerate operators in Liouville theory, i.e. $A_1$ conformal Toda field theory. The curve on the Riemann surface  along which the monodromy is computed depends on the  charge of  the loop operator in the way described in \cite{Drukker:2009tz}.  In this  paper we focus on electrically charged loops, i.e. Wilson loops. The simplest example of such  an operator is given by a loop  that has fundamental charge respect to only  one  of  the gauge groups in the theory.  According to \cite{Drukker:2009id} \cite{Alday:2009fs} \cite{Drukker:2009tz}, the two dimensional representation of this Wilson loop is given by the monodromy of a chiral degenerate operator evaluated along a closed curve encircling  the tube  associated to the relevant gauge group, see Figure \ref{figure1}.

The prescription for computing the monodromy  around this curve was given in \cite{Drukker:2009id} \cite{Alday:2009fs} in terms of   fusion moves  and  braiding moves. In the  following we review the prescription in a  way  that can be  easily  generalizated   to the  $SU(N)$ case. 

\begin{figure}[h]
\center{\includegraphics[width=10cm]{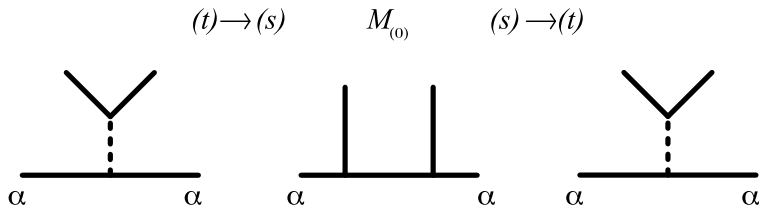}}
\caption{The prescription to compute the Wilson loop include a change  of  basis from the $t$-channel to the $s$-channel, a monodromy around  $z=0$ and finally  a change  of  basis from $s$-channel to the $t$-channel.  Dashed lines represent identity states.}\label{figure2}
\end{figure}

First,  one   inserts in the correlator two  degenerate chiral fields $V_{-\frac{b}{2}}(z)$, $V_{-\frac{b}{2}}(1)$  and  fuse them to the identity.  The starting point is thus the conformal block in the $t$-channel  with the identity operator as internal state,  we denote it as  ${\cal F}^{(t)}_2$.\footnote{Since we are now  looking at the $A_1$ Toda (Liouville) theory, there are  2  conformal blocks  in  the   $t$-channel but only one  of  them has the identity  as   internal state.  In the  following, we will consider $A_{N-1}$ Toda  and the conformal block with  the  identity as internal state  will be  denoted as ${\cal F}^{(t)}_N$.}    Then,     ${\cal F}^{(t)}_2$ is written in terms of the conformal blocks in the $s$-channel  ${\cal F}^{(s)}_k$ using the inverse fusion matrix  $F^{-1}_{kr}$, i.e.    ${\cal F}^{(t)}_2=F^{-1}_{2r}{\cal F}^{(s)}_r$.  The $F^{-1}$ describes  the  linear  relation between two  different  basis  of the   hypergeometric  functions \bref{hsol}. In particular $\Lambda^{(t)}=F^{-1}\Lambda^{(s)}$  and $\Lambda^{(s)}=F\Lambda^{(t)}$.   In the next step, the degenerate operator $V_{-\frac{b}{2}}(z)$ is moved around the  operator $V_\alpha(0)$.  This represents a monodromy around  the $z=0$ singular point. In the $s$-channel this monodromy acts diagonally on the conformal blocks and is given by a  diagonal matrix $e^{2\pi i b\alpha}M^{(s)}_{(0)kr}$.  Note that  $M^{(s)}_{(0)kr}$ is the monodromy associated to hypergeometric functions  $\Lambda^{(s)}_k$ and $e^{2\pi i b\alpha}$ is the monodromy of the factor $z^{b\langle \alpha, h_1 \rangle}(1-z)^{\frac{-b^2}{N}}$.  The result is then rotated back to the  $t$-channel using the fusion matrices $F_{kr}$  and  projected on the ${\cal F}^{t}_2$ conformal block, see Figure \ref{figure2}. 

More precisely, denoting with $\hat{{\cal L}}$ the set of operations  we have  just  described, we have
\begin{equation}\label{wl}
\hat{{\cal L}}\cdot{\cal F}_2^{(t)}= e^{2\pi i b\alpha} F^{-1}_{2k}M_{(0)kr}^{(s)}F_{r2}{\cal F}_2^{(t)}
\end{equation}
and thus the conformal block is  an eigenstate of $\hat{{\cal L}}$. It  follows  that the Wilson  loop is  described inserting in the   correlator
\begin{equation}
{\cal L}= e^{2\pi i b\alpha}F^{-1}_{2k}M_{(0)kr}^{(s)}F_{r2}=e^{2\pi i b\alpha}M_{(0)22}^{(t)}.
\end{equation}
The Wilson loop operator is thus given by the $(2,2)$ component of the hypergeometric monodromy matrix around $z=0$ expressed in the $t$-channel basis multiplied by the monodromy of the factor $z^{b\langle \alpha_1, h_1 \rangle}(1-z)^{\frac{-b^2}{N}}$ that appears in   \bref{cblock}. We now proceed to generalize this result to the $A_{N-1}$ Toda field theory.  We are  interested in Wilson loops in the  fundamental representation, we propose that these operators  are  associated  to the  monodromy of a degenerate   field $V_{-b\omega_1}$.  Since $\omega_1$ is the  first of the fundamental weights, this operator is naturally associated to the fundamental representation of $SU(N)$. Generalizing the procedure in  \cite{Drukker:2009id} \cite{Alday:2009fs}, besides  $V_{-b\omega_1}(z)$ we  introduce also the degenerate field $V_{-b\omega_{N-1}}(1)$ and  consider  the conformal block where the two  fields fuse  to the  identity.  Indeed it results that 
\bea\label{opefund}
V_{-b\omega_1}\cdot V_{-b\omega_{N-1}}=[V_{0}]+[V_{-b(\omega_1+\omega_{N-1})}]
\eea
and the identity operator is  included  in the OPE of the two  fields.  The   channel  where the two degenerate  fields are  fused each other is  associated   to the conformal blocks  ${\cal F}^{(t)}_k=z^{b\langle \alpha_1, h_1 \rangle}(1-z)^{\frac{-b^2}{N}}\Lambda^{(t)}_k$ defined in the previous section. In order  to  understand  which one  of the conformal blocks describes  the state  where the degenerate  fields are fused to the identity,  we compute the monodromy  around  the $z=1$ singularity. The  monodromy is  computed  considering the  contribution of the  of the $\Lambda^{(t)}_k$  solutions and  the  contribution of the $z^{b\langle \alpha_1, h_1 \rangle}(1-z)^{\frac{-b^2}{N}}$ factor. It results ${\cal F}^{(t)}\longrightarrow e^{-2\pi i\frac{b^2}{N}} M_{(1)}^{(t)}{\cal F}^{(t)}$ where $M_{(1)}^{(t)}$ is the  monodromy matrix around $z=1$ of the solutions $\Lambda^{(t)}_k$. More explicitly 
\bea
 \left( \begin{array}{c}
{\cal F}^{(t)}_1\\
{\cal F}^{(t)}_2\\
.  \\
.\\
{\cal F}^{(t)}_N \\
\end{array} \right)\longrightarrow
 \left( \begin{array}{ccccc}
\left(B_{-b(\omega_1+\omega_{N-1})}^{-b\omega_1,-b\omega_{N-1}}\right)^2&0& & &0\\
0&\left(B_{-b(\omega_1+\omega_{N-1})}^{-b\omega_1,-b\omega_{N-1}}\right)^2& & &0\\
. &.& & &. \\
. &.& & &.\\
0&.& & &\left(B_{0}^{-b\omega_1,-b\omega_{N-1}}\right)^2\\
\end{array} \right)
 \left( \begin{array}{c}
{\cal F}^{(t)}_1\\
{\cal F}^{(t)}_2\\
.  \\
.\\
{\cal F}^{(t)}_N \\
\end{array} \right)
\eea
where 
\begin{equation}\label{braid}
B_{\alpha_1}^{\alpha_2,\alpha_3}
=e^{i\pi(\Delta(\alpha_1)-\Delta(\alpha_2)-\Delta(\alpha_3))}
\end{equation}
and  $\Delta(\alpha)=\frac{1}{2}\langle \alpha, 2 Q -\alpha \rangle$ is the conformal dimension of the Toda primary operators. 
We thus  identify ${\cal F}^{(t)}_N$ as  the  conformal block with  the  identity in the  internal channel, while ${\cal F}^{(t)}_1,\ldots,{\cal F}^{(t)}_{N-1}$ are the conformal blocks  with the internal field in the adjoint representation, in agreement with \bref{opefund}.

Generalizing straightforwardly the prescription of  \cite{Drukker:2009id} \cite{Alday:2009fs}, we have that  an $SU(N)$ Wilson loop in the  fundamental representation is  associated to  
\begin{equation}
{\cal L}=e^{2\pi i b\langle \alpha, h_1 \rangle} M_{(0)NN}^{(t)},
\end{equation}
where $M_{(0)NN}^{(t)}$ is the $(N,N)$ component of the monodromy matrix around $z=0$, expressed in the $t$-channel basis. To compute this  quantity, it is possible to  use the matrices that relate the different set  of  solutions \bref{hsol}\footnote{See  for  instance \cite{norlund}.}. We will take  however a different route, using properties of  the  monodromy group of the  generalized hypergeometric  equation.

\begin{figure}[h]
\center{\includegraphics[width=10cm]{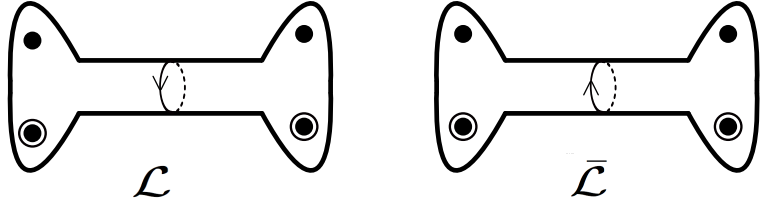}}
\caption{A Wilson loop in the fundamental representation  ${\cal L}$ and  a  Wilson loop in the  anti-fundamental representations $\bar{\cal L}$ are associated to curves with opposite orientation. In this example  we consider the gauge  theory  associated to the sphere with two simple  punctures and two full punctures.}\label{figure3}
\end{figure}

We have  already  mentioned  that  the  monodromy  group is  invariant under conjugation.  An element  in the conjugacy  class   corresponds  to  a  particular  set  of  solutions and  through conjugation, it  is  possible  to  know the  form  of the  monodromy  matrices  for  other  sets of  solutions.  
An explicit  realization of the monodromy  group for the generalized hypergeometric  equation was  given in \cite{BH}  \cite{O}, we review this construction in Appendix B.  Considering  a  certain conjugation, it  is  possible  to  obtain  the  monodromy  matrices  for  the basis of  functions $\Lambda^{(1)}$ where  $M_{(1)}^{(t)}$ is  diagonal. In this  basis it results\footnote{See Appendix B for  details.} 
\bea
M_{(0)NN}^{(t)}=e^{-2\pi i b\langle \alpha, h_1 \rangle} \frac{e^{i\pi bq}-e^{-i\pi bq}}{e^{i\pi bqN}-e^{-i\pi bqN}}\sum_k e^{2\pi i b\langle \alpha - Q, h_k \rangle}
\eea
where $h_k$ are  the $N$ weights of the  fundamental representation of $SU(N)$ and $q=\left(b+\frac{1}{b}\right)$.  Following \cite{Wyllard:2009hg}
we take $\alpha=\tilde{a}+Q$ where $\tilde{a}$ is an imaginary  $(N-1)$ vector  that parameterizes the   Cartan  subalgebra  of  $SU(N)$. Expanding $\tilde{a}$ on the  basis  of  simple roots as $\tilde{a}=\sum_{i=1}^{N-1}a_ie_i$ we  get   
\bea\label{wlta}
\sum_k e^{2\pi i b\langle \alpha - Q, h_k \rangle}&=&\left(e^{i2\pi b a_1 }+e^{i2\pi b (a_2-a_1)}+\ldots+e^{i2\pi b(a_{N-1}-a_{N-2})}+e^{i2\pi b(-a_{N-1})}\right)\nonumber\\
&=& \hbox{Tr}_{F}\, e^{i2\pi b a }
\eea
where  $a$ is an anti-hermitian traceless $N\times N$  matrix  and the trace is  evaluated in the  fundamental representation $F$.    We thus  conclude  that  a Wilson loop in the  fundamental representation of the gauge  group $SU(N)$, in the dual $A_{N-1}$ Toda theory is  described by  
\begin{equation}
{\cal L}=\frac{1}{[N]_{e^{i\pi bq}}}\hbox{Tr}_{F}\,e^{i2\pi b a},
\end{equation} 
where $[N]_{e^{i\pi bq}}=\frac{e^{i\pi bqN}-e^{-i\pi bqN}}{e^{i\pi bq}-e^{-i\pi bq}}$ is the quantum deformed number $N$ where the  parameter  of the  deformation is $e^{i\pi bq}$.  It  follows  immediately that in the  limit $b\rightarrow 1$,  the   Wilson loop reduces  to 
\begin{equation}
{\cal L}=\frac{1}{N}\hbox{Tr}_{F}\, e^{i2\pi  a },
\end{equation} 
in agreement with the result of Pestun \bref{wlpest}.  Considering $N=2$, we have the Wilson loop in $A_1$ Toda theory, i.e. Liouville  theory. It results \footnote{In this formula $a$ is one of the  eigenvalues of the traceless anti-hermitian matrix.}  
\bea
{\cal L}=\frac{\cos{(2\pi b a)}}{\cos{(\pi b q)}}  
\eea
in agreement with \cite{Drukker:2009id} \cite{Alday:2009fs}.\footnote{In \cite{Alday:2009fs}, the authors  consider  a different  normalization for loop operators.} Repeating the same procedure considering  the inverse matrix $M_{(0)}^{-1}$ instead of $M_{(0)}$, we compute the loop  operator  associated to the monodromy  defined around  the  same  curve but with  opposite  orientation\footnote{See Appendix B for details.}.  We denote this operator  with $\bar{{\cal L}}$. It results
\begin{equation}
\bar{{\cal L}}=\frac{1}{[N]_{e^{i\pi bq}}}\hbox{Tr}_{\bar{F}}\, e^{i2\pi b a},
\end{equation} 
where the trace is  evaluated in  the anti-fundamantal representation $\bar{F}$. We conclude  that in the $N>2$ case, to  completely  characterize a monodromy  operator, it  is  necessary to specify  also the  orientation of  the  curve  along  which the  monodromy  is  evaluated, see Figure \ref{figure3}.

\subsection*{Acknowledgements}
We would like to thank Nadav Drukker,  Jaume Gomis, Sara Pasquetti and Donovan Young for  useful discussions. This work  has been 
supported by the Volkswagen Foundation.

\appendix

\section{Some Lie Algebra Notions}
Given  a  finite  dimensional  semisimple Lie  algebra with rank $r$ (see for  instance \cite{Humph}), we  denote  $H^i$  $(i=1,\ldots, r)$ the elements in the Cartan subalgebra  and   $E^\alpha$ the remaining operators.  It is possible to consider a basis  of  operators  such that 
\begin{equation}
[H^i,E^\alpha]=\alpha^iE^\alpha\qquad\qquad [E^\alpha,E^{-\alpha}]=\frac{2}{\alpha\cdot\alpha}\alpha\cdot H 
\end{equation}
where  the $r$-dimensional vectors  $\alpha$ are  the  roots of the algebra and the scalar  product is    $\alpha\cdot\beta\equiv\sum_{i=1}^{r}\alpha^i\beta^i$. It is useful to define  the root  space   product  $\langle \beta,\alpha\rangle$ as
\begin{equation}
\langle \beta,\alpha\rangle\equiv\frac{2}{\alpha\cdot\alpha}\beta\cdot\alpha
\end{equation}
We  denote  with $\Phi$  the  set  of   roots  of  the  algebra.  The   simple roots $e_i$  $(i=1,\ldots, r)$ form  a  subset of roots  $\Delta\subset\Phi$ that are a basis for  the full root  space. It results that  any  root  in $\Phi$ can be  written as linear  combination of  simple  roots, i.e. $\alpha=\sum_{i=1}^{r}k_i e_i$ where  $k_i$ are  all  positive  integers or  all  negative  integers.  The  matrix defined  by  $\langle e_i,e_j \rangle=K_{ij}$ is the Cartan matrix  and its  entries are  integer  numbers.
For an  arbitrary representation of the algebra, it is possible to define a basis $|\lambda\rangle$ such  that $H^i|\lambda\rangle=\lambda^i |\lambda\rangle$. The $r$-dimensional vectors $\lambda$  are called weights  and satisfy the relation $\langle \lambda,\alpha\rangle\in  \mathbb{Z}$. They can be   expanded  on a  basis of  fundametal  weights $\omega_i$ such  that $\lambda=\sum_{i=1}^{r}\delta_i \omega_i$  and the coefficient of the expansion $\delta_i$ are integer  numbers called Dynkin labels. The fundamental weights are defined such that    $\langle\omega_i, e_j\rangle=\delta_{ij}$.  The highest weight of a  representation is the weight  for which the sum of the Dynkin labels is the highest. It can be shown that the highest weight completely characterize the representation.   Roots and simple roots are the weights and  fundamental weights of the adjoint representation. The  Weyl vector $\rho$ is defined as the  sum of all the fundamental weights,  i.e. $\rho=\sum_{i=1}^r\omega_r$. It follows  that $\langle\rho, e_i\rangle=1$.

\subsection{$ A_{N-1}$ Algebra}
A linear  representation of  the  $A_{N-1}$  algebra is  given by $N\times N$ traceless  matrices. Denoting  as $a_{ \mu  \nu}$ the matrix that have  entry $1$ at position $(\mu,\nu)$ and zero  elsewhere, we can construct the generators  $E^{\alpha_{\mu\nu}}=a_{\mu\nu}$ for $(\mu\neq \nu)$ and $H^i=\sum_{\mu=1}^{N}\epsilon^i_{\mu}a_{\mu\mu}$, where $\epsilon^{i}_{\mu}=(\epsilon_{1}^{i},\ldots, \epsilon_{N}^{i})$  satisfy $\sum_{\mu=1}^{N}\epsilon^{i}_{\mu}=0$. We have  
\begin{equation}\label{commusun}
[H^i,E^{\alpha_{\mu\nu}}]=(\epsilon_\mu^i-\epsilon_\nu^i)E^{\alpha_{\mu\nu}}
\end{equation}
thus  the  positive  roots  are  given by $\alpha^i_{\mu\nu}=(\epsilon_\mu^i-\epsilon_\nu^i)$ where $1<\mu<\nu<N$ and  the simple  roots  are
 $e_{j}=(\epsilon_j^i-\epsilon_{j+1}^i)$ where  $1<j<N-1$.  In the main text we omit indices labeling the components of  roots or weights. The Cartan matrix $K_{ij}=\langle e_i, e_j\rangle$ is given by 
 \begin{equation}\label{carmat}
     K_{ij}=\pmatrix{
    2  & -1 & 0 & \dots & 0&0\cr
    -1 &  \;2 & -1 & \ldots& 0&0\cr
    0 & -1 & 2&\ldots&0&0\cr
    \cdot& \cdot &\cdot&\cdot &\cdot&\cdot \cr
    0 & 0 & 0 & \ldots & 2&-1\cr
   0 & 0 & 0 & \ldots & -1&2}
\end{equation}
The fundamental representation $F$  of the $SU(N)$ algebra has the first fundamental weight as highest  weight, i.e.  $h_1=\omega_1$. The set of  $N$ weights of  $F$  is  given by 
\begin{equation}\label{fwei}
h_k=\omega_1-e_1-\ldots- e_{k-1}
\end{equation} 
where $k=1,\ldots N$ and we assumed $e_0=0$. The anti-fundamental representation $\bar{F}$  of the $SU(N)$ algebra has the last fundamental weight $\omega_{N-1}$ as highest  weight.  $\bar{h}_{k}$ are the   $N$ weights of  $\bar{F}$.  Simple roots  and  fundamental  weights  are  related by the Cartan matrix as $e_i=\sum_{j}K_{ij}\omega_j$. Other useful relations  are 
\bea
\langle \rho , h_1 \rangle=\frac{N-1}{2},\qquad
\langle h_1 , h_1 \rangle=\frac{N-1}{N},\qquad
\langle \rho , \sum_{k=1}^N h_k \rangle=0.\label{formul}
\eea

\section{The Hypergeometric Monodromy Group}

The monodromy group of the generalized hypergeometric equation \bref{hypeq}  was analyzed in details in \cite{BH} and \cite{O}. An explicit representation of the group is  given by 
\bea\label{mongroup}
M_{(\infty)}=A\qquad M_{(0)}=B^{-1}\qquad M_{(1)}=A^{-1}B
\eea     
where 
\bea
A=\left( \begin{array}{ccccc}
0 & 0 & \ldots & 0 & -c_{N} \\
1 & 0 & \ldots & 0 & -c_{N-1} \\
0 & 1 & \ldots & 0 & -c_{N-2} \\
 &  & \ldots &  &  \\
0 & 0 & \ldots & 1 & -c_1 \\
\end{array} \right),\qquad\qquad
B=\left( \begin{array}{ccccc}
0 & 0 & \ldots & 0 & -b_{N} \\
1 & 0 & \ldots & 0 & -b_{N-1} \\
0 & 1 & \ldots & 0 & -b_{N-2} \\
 &  & \ldots &  &  \\
0 & 0 & \ldots & 1 & -b_1 \\
\end{array} \right)
\eea
 and the entries $c_k$ and $b_k$ are defined by 
 \bea
 \prod_{k=1}^{N}(t-e^{2\pi i A_k})=t^N+c_1t^{N-1}+\ldots +c_N,\quad  \prod_{k=1}^{N}(t-e^{2\pi i B_k})=t^N+b_1t^{N-1}+\ldots +b_N
 \eea

The  representation of the monodromy  group given in  \bref{mongroup},  produces the monodromy matrices associated  to a certain basis of independent solutions.  Considering  a different  basis correspond to a  conjugation of the monodromy  matrices.  In particular, we  are  interested  in the  expression of $M_{(0)}$ and $M_{(0)}^{-1}$  in a  basis  where $M_{(1)}$ is  diagonal, i.e. in the basis of solutions $\Lambda^{(t)}$.  It results  that  given  a certain $D\in GL(N,\mathbb{C})$, we have 
\bea
M_{(1)}^{(t)}&=&D^{-1}{M}_{(1)}D=\left( \begin{array}{ccccc}
1 & 0 & \ldots & 0 & 0 \\
0 & 1 & \ldots & 0 & 0\\
 &  & \ldots &  &  \\
0 & 0 & \ldots & 1 & 0\\
0 & 0 & \ldots & 0 & \frac{b_N}{c_N} \\
\end{array} \right),\\ \nonumber \\ \nonumber  \\ 
M_{(0)NN}^{(t)}&=&D^{-1}_{Nk}{M}_{(0)kr}D_{rN}=\frac{-c_N b_{N-1}+b_N c_{N-1}}{(c_N-b_N)b_N}, \\  \nonumber  \\ 
M_{(0)NN}^{-1(t)}&=&D^{-1}_{Nk}{M}_{(0)kr}^{-1}D_{rN}=\frac{(-c_1 + b_1) b_N}{c_N - b_N}.
\eea
where $M_{(0)NN}^{(t)}$ and $M_{(0)NN}^{-1(t)}$  are the $(N,N)$ components of the matrices $M_{(0)}^{(t)}$ and  $M_{(0)}^{-1(t)}$.
The parameter $c_N$, $c_{N-1}$ and $c_1$ are  related to the $A_k$  parameters as
\bea
c_{N}&=& (-1)^N \prod_{k=1}^{N}e^{2\pi i A_k},\nonumber\\ c_{N-1}&=& (-1)^{N-1} \sum_{k_1>k_2>\ldots >k_{N-1}\geq 1}^{N}e^{2\pi i A_{k_1}}e^{2\pi i A_{k_2}}\ldots e^{2\pi i A_{k_{N-1}}},\nonumber\\
c_{1}&=&-\sum_k e^{2\pi i A_k}
\eea 
and in analogous way   $b_N$, $b_{N-1}$ and $b_{1}$ are related to $B_k$.  Expressing   $A_k$ and $B_k$ in terms of the  Toda momenta using \bref{momenta} and considering $\alpha_1=\alpha$ and $\alpha_2=2Q-\alpha$, applying  formulas like the \bref{formul},  we have   
\bea
M_{(1)}^{(t)}&=&D^{-1}{M}_{(1)}D=\left( \begin{array}{ccccc}
1 & 0 & \ldots & 0 & 0 \\
0 & 1 & \ldots & 0 & 0\\
 &  & \ldots &  &  \\
0 & 0 & \ldots & 1 & 0\\
0 & 0 & \ldots & 0 & e^{2\pi i b^2N}\\
\end{array} \right),\\ \nonumber \\
M_{(0)NN}^{(t)}&=&e^{-2\pi i b\langle \alpha, h_1 \rangle} \frac{e^{i\pi bq}-e^{-i\pi bq}}{e^{i\pi bqN}-e^{-i\pi bqN}}\sum_k e^{2\pi i b\langle \alpha - Q, h_k \rangle},\\   \nonumber\\
M_{(0)NN}^{-1(t)}&=&e^{2\pi i b\langle \alpha, h_1 \rangle} \frac{e^{i\pi bq}-e^{-i\pi bq}}{e^{i\pi bqN}-e^{-i\pi bqN}}\sum_k e^{-2\pi i b\langle \alpha - Q, h_k \rangle}.
\eea

\providecommand{\href}[2]{#2}\begingroup\raggedright\endgroup

\end{document}